\documentclass[12pt]{article}
\def\singlespacing{\baselineskip=12pt}

\begin{document}
\input{epsf}
\singlespacing
%\doublespacing

\title{\bf{\Large
{\bf Nonlinear limits to the information capacity of 
optical fiber communications}}}
\author{
Partha P Mitra\\
Jason B Stark\\
Bell Laboratories,  Lucent Technologies, Murray Hill, NJ 07974\\
}
\maketitle
 
\bigskip

{\bf 
The exponential growth in the rate at which information can be 
communicated through an optical fiber is a key element in the 
so called information revolution. However, like all exponential 
growth laws, there are physical limits to be considered. 
The nonlinear nature of the propagation of light in optical fiber
has made these limits difficult to elucidate. Here we obtain 
basic insights into the limits to the information capacity 
of an optical fiber arising from these nonlinearities. 
The key simplification lies in relating the nonlinear channel  
to a linear channel with multiplicative noise, for which 
we are able to obtain analytical results. In fundamental 
distinction to the linear additive noise case, the capacity does 
not grow indefinitely with increasing signal power, but has a 
maximal value. The ideas presented here have broader implications 
for other nonlinear information channels, such as those involved 
in sensory transduction in neurobiology. These have been often 
examined using additive noise linear channel models, and as 
we show here, nonlinearities can change the picture qualitatively.   
}

The classical theory of communications {\it \`{a} la} Shannon
\cite{shannon} was developed mostly in the context of 
linear channels with additive noise, which was adequate for 
electromagnetic propagation through wires and cables that 
have until recently been the main 
conduits for information flow. Fading channels 
or channels with multiplicative noise have been considered, 
for example in the context of wireless communications 
\cite{wireless}, 
although such channels remain theoretically less tractable 
than the additive noise channels. However, with the advent 
of optical fiber 
communications we are faced with a nonlinear propagation 
channel that poses major challenges to our understanding. 
The difficulty resides in the fact that the input output 
relationship of an optical fiber channel is obtained by 
integrating a nonlinear partial differential equation and 
may not be represented by an instantaneous nonlinearity.  
Channels where the nonlinearities in the input output 
relationship are not instantaneous are  
in general ill understood, the optical fiber simply being 
a case of current relevance. The understanding of such 
nonlinear channels with memory are of fundamental interest, 
both because communication rates through optical fiber are 
increasing exponentially and we need to know where the 
limits are, and also because understanding such channels may 
give us insight elsewhere, such as into the design 
principles of neurobiological information channels at the 
sensory periphery. 

The capacity of a communication channel is the maximal rate
at which information may be transferred through the channel 
without error. The capacity can be written as a product of 
two conceptually distinct quantities, the spectral bandwidth
$W$ and the maximal spectral efficiency which we will 
denote ${\cal C}$. In the classic capacity formula 
for the additive white Gaussian noise channel with an average
power constraint, $C = W \log(1+S/N)$ \cite{shannon}, 
the spectral bandwidth $W$, which has dimensions of inverse time, 
multiplies the dimensionless maximal spectral efficiency 
${\cal C} = \log(1+S/N)$. Here $S$ and $N$ are the signal and 
noise powers respectively. It is instructive to examine 
this formula in the context of an optical fiber. Since 
the maximal spectral efficiency is logarithmic in the signal to 
noise ratio (SNR), it can never be too large in a realistic 
situation, so that
the capacity is principally determined by the bandwidth $W$. 
In the case of an optical fiber, the intrinsic loss mechanisms 
of light propagating through silica fundamentally limits $W$ to 
a maximum of about $50 THz$ \cite{slusher} corresponding to a 
wavelength range of about $400 nm ~ (1.2-1.6 \mu)$. This is 
to be compared with current systems where the total bandwidth 
is limited to about $15 ~ THz$. If the channel was linear, 
the maximal spectral efficiency would be ${\cal C} = \log(1+S/N)$, 
$S$ being input light intensity and $N$ the intensity 
of amplified spontaneous emission noise in the system. 
An output SNR of say 100 (i.e. 20dB), would then yield  a
spectral efficiency of 6.6, which for a $50THz$ channel  
would correspond to a capacity of $330 ~ Tbit/sec$. The channel, 
of course, is not linear; how do the nonlinearities impact the spectral 
efficiency of the fiber? The basic conclusion of the present 
work is that the impact is severe and qualitative. As shown 
in Fig.1, the effect is a saturation and eventual decline of
spectral efficiency as a function of input signal power, in 
complete contrast with the linear channel case. We now proceed 
to motivate and discuss this result. 

It is widely recognised that nonlinearities impair the channel 
capacity. However, estimation of the impact of the nonlinearities 
on channel capacity has remained {\it ad hoc} from an information 
theory perspective. Here we obtain what appears to be the first 
systematic estimates (Fig.1) 
for the maximal spectral efficiency of an optical 
fiber channel as a function of the relevant parameters. In basic
distinction to the linear channel, our considerations indicate that 
the maximal spectral efficiency does not grow indefinitely with signal 
power, but reaches a maximum of several bits and eventually 
{\it declines}, as illustrated in Figure 1. 
It is to be noted that current systems use a 
binary signalling scheme which limits the 
achievable spectral efficiency {\it a priori} to $1$ bit, 
and to reach the higher spectral efficiencies predicted by the 
theory, multi-bit signalling schemes would have to be used. 
Since the spectral efficiencies of current systems are already
approaching $1$ bit, it is clear that 
the limits discussed here will be of 
practical relevance in the future. 

Although a number of nonlinearities are present in light propagation
in a fiber, we concentrate on the most important one for fiber 
communications, namely the dependence of the refractive index 
(and therefore the propagation velocity of light) 
on the light intensity, $n = n_0 + n_2 I$. This 
nonlinearity is weak, but its effects accumulate due to the long 
propagation distances involved in fibre communications, and is 
responsible for the effects considered here. Three principle physical 
parameters characterising the propagation are of interest: the group 
velocity dispersion $\beta\sim 10 ps^2/km$, the propagation loss 
$\alpha \sim 0.2~dB/km$ and the strength of the nonlinear refractive 
index, usually expressed in terms of the parameter 
$\gamma \sim 1/W/km$. The propagation loss is compensated by 
interposing optical amplifiers into the system. Each amplifier also 
injects spontaneous emission noise into the system with strength 
$I_1 = a G h \nu \Delta \nu$ \cite{agrawal1}, 
with $G$ being the amplifier gain, 
$h$ the Planck's constant, $\nu$ and $\Delta\nu$ being the centre
frequency and frequency bandwidth of light respectively. Here 
`$a$' is a numerical constant (which we assume to be 2). For 
$n_s$ spans of fiber interspersed with amplifiers that make the 
total channel gain unity, the effects of absorption may be 
accounted for simply by redefining the system length in terms 
of an effective length, $L_{eff} \sim n_s/\alpha$. If the nonlinearity
were absent ($\gamma = 0$), we would have obtained, for the 
maximal spectral efficiency, ${\cal C}_0 = \log(1+I/I_n)$, 
$I$ being the input power and $I_n = n_s I_1$ being the total 
additive noise power. Note that ${\cal C}_0$ declines 
logarithmically with system length, and would eventually 
vanish for infinitely long systems. Note also that although 
spectral efficiency is dimensionless, it is often written 
for convenience with the ``units''  bits/sec/Hz. 

For a variety of reasons, the principal one being limitations 
in the electronic bandwidth, it is impractical to modulate 
the full optical bandwidth at once. Instead, current attempts 
towards achieving maximal information throughput involve so called 
Wavelength Division Multiplexing (WDM) \cite{slusher}, where the whole optical 
bandwidth is broken up into disjoint frequency bands (``channels'')
each of which is modulated separately. We confine our attention 
to such systems (which from an information theory perspective 
corresponds to the ``multi-user'' case) \cite{coverthomas}, though we also 
comment on the ideal case of utilising the full optical bandwidth 
for a single data stream (the ``single user'' case). Quantitatively, 
the single user case is expected to have larger maximal spectral 
efficiencies, though we will argue that it shows the same 
qualitative behaviour as the multi-user case. The difference 
between the two reside in the fact that in the multi-user case, 
each channel is an independent information stream, and appears 
as an additional source of noise to every other channel due 
to nonlinear mixing. 

The nonlinear propagation effects in the evolution of the 
electric field amplitude involve a cubic term in the electric
field. In a WDM system, the nonlinearities are classified by 
the field amplitudes participating in this cubic term for the 
evolution of the field amplitude of a given channel: 
self phase modulation denotes the case where all three fields 
belong to the same channel, cross phase modulation where two fields
belong to a different channel and one to the same channel, and 
four wave mixing denotes the case where all three amplitudes 
belong to different channels. Out of these terms, four wave 
mixing gives rise to additive noise to the channel of interest
and will not be considered further in this paper. One reason for
this is that four wave mixing is strongly suppressed by dispersion 
when the channel spacings are substantial. Its effects can be 
accounted for by augmenting the additive noise term in the subsequent 
considerations. We also neglect self phase modulation effects, 
since these effects are deterministic for the given channel and in 
principle could be reduced by using nonlinear precompensation. 
Finally, we are left with cross phase modulation, which appears 
to be the principle source of nonlinear capacity impairment in 
the multiuser case for realistic parameter ranges. A further 
reason for our focus on cross phase modulation is that 
it gives rise to multiplicative noise, which gives rise to 
qualitatively new effects in the channel capacity. 

We model the propagation channel in the presence of cross phase 
modulation by means of a {\it linear Schroedinger equation} with 
a {\it random potential} fluctuating both in space and time. This 
is easily justified starting from the nonlinear Schroedinger
equation description commonly used to describe light propagation 
in single mode optical fibres \cite{agrawal}.   
Cross phase modulation arises from terms in the equation where 
the field intensity in the nonlinear refractive index 
is approximated by the sum of the field intensities in the channels 
other than the one for which the propagation is being studied. 
Therefore, if only cross phase modulation effects were retained, 
the propagation equation for the field amplitude in channel $i$ 
then becomes 
\begin{equation} 
i \partial_z E_i = {\beta_2 \over 2} \partial_t^2 E_i +V(z,t) E_i,
\label{nlse}
\end{equation} 
where $V(z,t) = - 2 \gamma \sum_{j\neq i} |E_j(z,t)|^2$, 
the sum being taken 
over the other channels. Since independent streams of information 
are transmitted in the other channels, $V(z,t)$ appears as a random 
noise term. Notice that the nonlinear propagation equation has 
now been reduced to a linear Schroedinger equation with a stochastic 
potential, so that the nonlinear channel has become a channel with 
multiplicative noise. We now need an adequate model for the stochastic
properties of $V(z,t)$. If the dispersion is substantial, we propose 
that $V(z,t)$ may be approximated by a Gaussian stochastic process 
short range correlated in both space and time. 
Since V is obtained by adding a 
large number of different channels, each of which is short range 
correlated in time ($\tau \sim 1/B$, where $B$ is the channel bandwidth),
we can expect V to have a correlation time of approximately $1/B$.
Dispersion causes the channels to travel at different speeds, thus 
causing $V$ to be short range correlated in space as well, with a 
correlation length related to the dispersion length. Since $V$ is 
a sum of intensities, it has nonzero mean, so we define 
$\delta V(z,t) = V(z,t) - \langle V \rangle $, where $\langle V \rangle $ denotes the average value 
of V. Removing a constant from the potential causes an overall 
phase shift independent of space and time, which is irrelevant to 
the present considerations.

The parameter of interest in the following is the integrated 
strength of the fluctuating field, 
$\eta = \int dz \langle \delta V(z,0) \delta V(0,0) \rangle$. 
In order to estimate $\eta$, we consider a simplified
propagation model for the channels other than the one 
of interest, in which nonlinearities are neglected, and 
stochastic bit streams at the inputs to the channels are 
propagated forward with constant group velocities. 
The group velocity difference between two channels separated
by a spacing $\Delta\lambda$ is $D\Delta\lambda$.
In this model with $n_c$ other channels evenly spaced 
by $\Delta\lambda$ around the channel of interest, 
each with intensity $I$ and bandwidth $B$, we obtain 
$\eta = 2 \ln(n_c/2) (\gamma I) ^2 / (B D \Delta \lambda)$.
Here $D$ is the dispersion parameter $D = -2\pi c\beta/\lambda^2$. 
Although this is a simplified model for the other channels, 
numerical simulations of propagation including the nonlinearities 
and dispersion for the side channels show that the estimate of 
$\eta$ is accurate. 

Note that the denominator in the expression of $\eta$ is the 
inverse of the dispersion 
length $L_D$ for the given channel spacing. This 
form for $\eta$ follows from assuming that $L_{eff} >> L_D$, 
since in this limit the integral defining $\eta$ is cut off
by $L_D$. If on the other hand, $L_{eff} \leq L_D$, the 
integral would be cut off by $L_{eff}$, so that one would 
have to replace $L_D$ by $L_{eff}$  in the equation for $\eta$. 
The fluctuation strength
scales with the logarithm of the number of channels rather 
than the total number since channels at larger 
spacings are suppressed proportionately to channel spacing. 
This suppression due to dispersion leads to the logarithmic factor
via a sum of the form 
$\sum_j 1/(\Delta \lambda_j) \propto \sum_j 1/j$. 

Within the model under consideration, the propagation down 
the fiber is given in terms of a propagator $U(t,t^{\prime};L)$
obtained by integrating the stochastic Schroedinger equation. 
For simplicity, we model the amplifier noise as an additive 
term with strength $I_n$ as defined earlier. The channel is specified
in terms of a relation between the input and output electric 
field amplitudes, 
$E_{out}(t) = \int dt^{\prime}U(t,t^{\prime};L)E_{in}(t^{\prime}) +n(t)$.
Since $U$ is stochastic, due to the underlying stochasticity of 
$V(z,t)$, the model corresponds to a channel with multiplicative noise. 
It is still intractable in terms of an exact capacity computation, 
but an analytic lower bound may now be obtained. This bound is based 
on the following information theoretic result (E.Telatar, private 
communications): the capacity ${\cal C}$ of a channel with input 
$X$ and output $Y$ related by a conditional distribution $p(Y|X)$ 
and an input power constraint $E(||X||^2)=P$ satisfies the inequalities
${\cal C} = max_{p(X)} I(X,Y) \geq I(X_G,Y) \geq I(X_G,Y_G) $
Here $I(X_G,Y)$ is the mutual information when $p(X)$ is chosen to 
be $p_G(X)$, a Gaussian satisfying the power constraint; $I(X_G,Y_G)$ 
is the mutual information of a pair $(X_G,Y_G)$ with the same 
second moments as the pair $(X,Y)$. The first inequality is trivial
since $p_G(X)$ is not necessarily the optimal input distribution. 
A proof of the second inequality is outlined in the methods section. 

The quantity $I(X_G,Y_G)$ for the channel defined above may be 
computed from knowledge of the correlators 
$\langle E_{in}(t) E^*_{in}(t^{\prime}) \rangle $, 
$\langle E_{out}(t) E^*_{out}(t^{\prime}) \rangle $ and 
$\langle E_{out}(t) E^*_{in}(t^{\prime}) \rangle $. 
The first is defined {\it a priori} through the assumption of 
bandlimited Gaussian white noise input with a power constraint. 
The second follows from the first using the unitarity of $U$. 
The third correlator requires computation of the average 
propagator $\langle U \rangle $, where the average is over realisations of 
$V(z,t)$. For a Gaussian, delta-correlated $V$, we obtain 
$\langle U(t,t^{\prime};L) \rangle  = \exp(-\eta L/2) U_0(t-t^{\prime};L)$ 
(see methods), where $U_0$ is the propagator for $V=0$. 
Assembling these results, we finally obtain an analytic expression
for a lower bound $C_{LB}$ to the channel capacity of the 
stochastic Schroedinger equation model: 
\begin{equation} 
C_{LB} = n_c B \ln (1+{e^{- ({I\over I_0}) ^2  } I \over I_n + 
(1-e^{- ({I\over I_0}) ^2}) I} )
\label{fibercap}
\end{equation}
where $I_0$ is given by 
\begin{equation}
I_0 = \sqrt{ {B D \Delta \lambda \over 2 \gamma^2 \ln(n_c/2) L_{eff} } }.
\label{limint}
\end{equation}

The fundamental departure from a linear channel in the 
above capacity expression is the appearance of an intensity scale 
$I_0$, which governs the onset of nonlinear effects.
To obtain an idea about the value of $I_0$, consider the parameter 
values $B=40 GHz$, $D = 20 ps/nm/km$, $\Delta \lambda = 1nm$, 
$\gamma = 1/W/km$, $n_c = 100$, 
$L_{eff} \approx  n_{s}/\alpha = 100km$. Then $I_0 =32 mW$.  
Examination of Eq.\ref{limint} shows 
that the intensity scale $I_0$ at which nonlinearities set in
shows reasonable dependence on all 
relevant parameters, namely it increases with increases in the 
dispersion, the bandwidth and the channel spacing, but decreases 
with increasing system length and number of channels. 

The most striking feature of Eq.\ref{fibercap} is that
instead of increasing logarithmically with signal intensity
like in the linear case, the capacity estimate actually peaks 
and then declines beyond a certain input intensity. 
From Eq.\ref{fibercap}, it is easily derived that the maximum 
value is given approximately by 
$C_{max} \approx {2\over 3} n_c B \ln(2 I_0/I_n)$, the 
maximum being achieved for an intensity 
$I_{max} \approx (I_0^2 I_n/2)^{1/3}$. The reason 
for this behaviour is that if we consider any particular channel, 
the signal in the other channels appear as noise in the channel 
of interest, due to the nonlinearities. This `noise' power increases
with the `signal' strength, thus causing degradation of the capacity
at large `signal' strength. The behaviour of Eq.\ref{fibercap} is 
graphically illustrated in Fig.1, where the spectral efficiency
(bits transmitted per second per unit bandwidth) is shown as 
a function of input power.

It is of interest to note that if the input intensity is kept 
fixed, the capacity bound declines exponentially with the system length. 
This is only to be expected, since the correlations of the electric field 
should decay exponentially due to the 
fluctuating potential in the propagation 
equation.  On the other hand, the maximal spectral efficiency given by 
$C_{max}$ declines only logarithmically in system length, 
in parallel with the behaviour for linear channels. It can therefore be inferred
 that if the input power was adjusted with system length instead of being 
kept fixed, the decline of spectral efficiency with system length 
will be logarithmic. 

Finally, we present qualitative arguments as to why the single user
case is expected to show the same non-monotonicity of spectral 
efficiency with the input signal intensities. In the multi-user case, 
the noise power as effectively generated by cross phase modulation 
grows as $I^3$ since it involves three signal photons. In the single 
user case, the cubic nonlinearity is a deterministic process that does 
not necessarily degrade channel capacity. However, subleading processes 
which involve two signal and one spontaneous noise photon still 
scale superlinearly in signal intensity, as $I^2 I_n$. 
Therefore, one should still observe the same behaviour of the effective
noise power overwhelming the signal at large signal intensities. Thus, 
we would still expect the spectral efficiency to decline at large 
input intensity, though not as rapidly in the multi-user (WDM) case.

\section*{Methods}

\subsubsection*{Gaussian bound to the channel capacity}
Proof of the inequality $I(X_G,Y) \geq I(X_G,Y_G)$:
define $p(X,Y)$ as the product $p_G(X) p(Y|X)$, and
$p_G(X,Y)$ to be the joint Gaussian distribution having
the same second moments as $p(X,Y)$. Also define $p_G(Y)$ 
to be the corresponding marginal of $p_G(X,Y)$. 
\begin{eqnarray} 
I(X_G,Y) & = & \int dX dY p(X,Y) 
\log({p(X,Y) \over p_G(X) p(Y)}) \\
& = &  \int dX dY p(X,Y) [ \log({p_G(X,Y) \over p_G(X) p_G(Y)}) 
- \log( {p_G(X,Y) \over p(X,Y)} {p_(Y)\over p_G(Y)} ) ] \\
\end{eqnarray}
Since $p(X,Y)$ and $p_G(X,Y)$ share second moments, the 
first term on the RHS is $I(X_G,Y_G)$. The second term 
may be simplified using the convexity of the logarithm, 
$\langle \log(f) \rangle  \leq \log(\langle f \rangle )$ to obtain 
\begin{eqnarray}
I(X_G,Y) & \geq & I(X_G,Y_G) -\log[  \int dX dY p_G(X,Y) 
{p(Y) \over p_G(Y)}] \\
& \geq & I(X_G,Y_G)
\end{eqnarray}
The second inequality follows by first performing the integral 
over $X$, and noting that \\
$\log(\int dY p(Y) )= \log(1) = 0 $. 

\subsubsection*{Derivation of the average propagator $\langle U \rangle $:} 
This can be done by resumming the perturbation series
exactly for $\langle U \rangle $, for delta correlated $V(z,t)$. Alternatively, 
in the path integral formalism \cite{feynmanhibbs}, 
\begin{equation}
\langle U(t,t^{\prime};L) \rangle  = U_0(t-t^{\prime};L) 
\langle \langle  \exp(i \int_0^L dz V(z,t(z))  \rangle  \rangle ,
\nonumber
\end{equation}
where the average is taken over $V$ as well as over paths $t(z)$ satisfying 
$t(0)=t$, $t(L)=t^{\prime}$. The result in the paper follows by  
performing the Gaussian average over $V$. Since 
$\phi = \int_0^L dz V(z,t(z))$ is a linear combination of 
Gaussian variables, it is also Gaussian distributed and 
satisfies $\langle \exp(i\phi) \rangle =\exp(-\langle \phi^2 \rangle /2)$. The result 
follows by noting that for delta correlated $V$, 
$\langle \phi^2 \rangle $ is a constant given by $\eta L$. The delta 
correlations need to be treated carefully, this can 
be done by smearing the delta functions slightly and leads
to the definition of $\eta$ given earlier in the paper.

\subsection*{Acknowledgements}
We gratefully acknowledge 
discussions with E.~Telatar, R.~Slusher, A.~Chraplyvy, G.~Foschini
and other members of the fiber capacity modelling group at Bell 
Laboratories. We would also like to thank D.~R.~Hamann and R.~Slusher 
for careful readings of the manuscript.

\subsection*{Figure Captions}

\begin{description}

\item [Figure 1.]

The curves in Fig.1 represent lower bounds to the spectral
efficiency for a homogeneous length of fiber for a multi-user 
WDM system, given analytically by Eq.2. Although the curves 
represent lower bounds, we argue in the text that the true 
capacity shows the same qualitative non-monotonic behaviour
with respect to input signal powers. The spectral efficiencies
displayed in the figure correspond to the capacity per unit 
bandwidth, ${\cal C} = C/(n \delta \nu)$. Here $\delta \nu$ 
includes both the channel bandwidths and the inter-channel 
spacing. The parameters used for the figure are $n_c = 100$, 
$L_{eff} = 100 km$, $D = 20 ps/nm/km$, $\delta \nu = 1.5B$
where $B=10GHz$ is the individual channel width. 
The two continuous curves
 correspond to $\gamma = 1/W/km$ and $\gamma = 0.1/W/km$, 
the lower curve corresponding to $\gamma=1$. 
The spontaneous noise strength $I_n$ is computed from the formula 
$I_n = a G h \nu B$ as explained in the text, with $a=2$, $G=1000$, 
$\nu=200THz$. The dotted 
curve represents the spectral efficiencies of the corresponding 
linear channels given by $\gamma = 0$. 

\end{description}


\begin{thebibliography}{99}

\bibitem{shannon} 
Shannon, C.~E.  A mathematical theory of communications., 
{\it Bell Syst.~Tech.~J.},  {\bf 27}, p.~379-423, 
p.~623-656 (1978).

\bibitem{wireless} 
Biglieri E., Proakis, J. \& Shamai S.,
Fading channels: Information-theoretic and communications aspects., 
{\it Information Theory Transactions} {\bf 44}:6 p.~2619-2692 (1998).  

\bibitem{slusher}
Glass, A.M. {\it et al.}, Advances in Fiber Optics. 
{\it Bell Labs Technical Journal} {\bf 5}, p. 168 (2000).

\bibitem{agrawal} 
Agrawal, G.~P., {\it Nonlinear Fiber Optics}, 
Academic Press, Inc., San Diego, 1995.  

\bibitem{agrawal1}
Agrawal, G.~P., {\it Fiber-Optic Communication Systems},
John Wiley \& Sons, Inc., New York, 1992, pp.~334.

\bibitem{coverthomas}
Cover, T.~M. \& Thomas, J.~A. {\it Information Theory},
John Wiley \& Sons, Inc., New York, 1991.

\bibitem{feynmanhibbs}
Feynman,  R.~P. \& Hibbs, R.~A., {\it Quantum Mechanics and Path Integrals},
McGraw-Hill, New York, 1965.

\end{thebibliography}
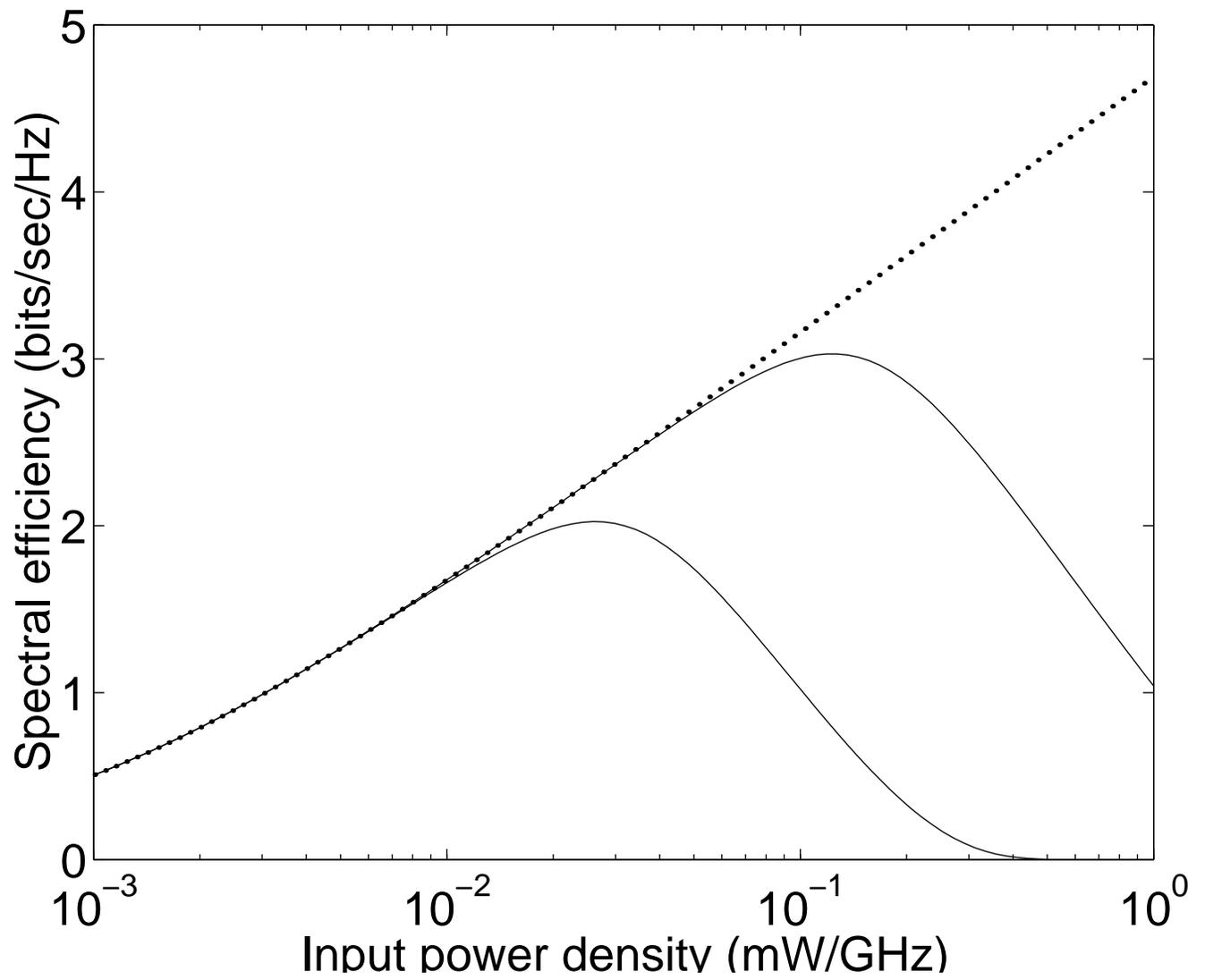
\end{document}